\documentclass[a4paper]{spie}  

\usepackage{amsmath,amsfonts,amssymb}
\usepackage{graphicx}
\usepackage[colorlinks=true, allcolors=blue]{hyperref}
\usepackage{aas_macros}

\title{Adaptive optics design status of MAORY, the MCAO system of European ELT}

\author[a]{Lorenzo Busoni}
\author[a]{Guido Agapito}
\author[a]{Cedric Plantet}
\author[c]{Sylvain Oberti}
\author[c]{Christophe Verinaud}
\author[c]{Miska Le Louarn}
\author[a]{Simone Esposito}
\author[b]{Paolo Ciliegi}
\affil[a]{INAF Osservatorio Astrofisico di Arcetri, Largo E. Fermi 5, Firenze, Italy}
\affil[b]{INAF Osservatorio di Astrofisica e Scienza dello Spazio di Bologna (OAS), via Gobetti 93/3, Bologna, Italy}
\affil[c]{European Southern Observatory (ESO),Karl-Schwarzschild-Str. 2, D-85748 Garching bei Muenchen, Germany}

\authorinfo{Send correspondence to L.B.: lorenzo.busoni@inaf.it}

\begin{document} 
\maketitle

\begin{abstract}
MAORY is the Multi-conjugate Adaptive Optics RelaY for the European ELT aimed at providing a 1 arcmin corrected field to MICADO, a near-infrared spectro-imager with a focus on astrometry.
In this paper we review the main requirements and analysis that justify the current adaptive optics architecture and subsystem requirements. We discuss the wavefront error budget allocation focusing on the worst offenders terms and on a statistical analysis of their dependence on atmospheric and sodium profiles.
We present an updated revision of the trade-off studies on the main AO parameters that, along with considerations coming from optical and mechanical subsystems, are used to define the preliminary design of the instrument. 
\end{abstract}


\section{INTRODUCTION}
\label{sec:intro}  
MAORY \cite{10.1117/12.2234585} is the Multi-Conjugate Adaptive Optics system of the ELT \cite{2018SPIE10702E..1PR}, providing large field diffraction limited correction for the near-infrared camera MICADO \cite{2018SPIE10702E..1SD}. MICADO's primary observing mode is astrometric imaging with field of 50"x50" and 19"x19" sampled at 4mas/px and 1.5mas/px respectively and wavelength coverage from 0.8 $\mu m$ to 2.4 $\mu m$. A Single-Conjugate Adaptive Optics system \cite{2018SPIE10703E..13C}, jointly developed by MAORY and MICADO, will also be available for compact target observations and for a phased AO integration at the ELT.

In this paper we report on the status of the MCAO system, currently in its preliminary design phase, focusing on the analysis of the requirements and interfaces that are driving the design of the adaptive optics configuration. 

\section{OVERVIEW OF MAORY REQUIREMENTS}
MAORY shall provide uniform correction over a 1 arcmin field delivering 30\% Strehl ratio (goal 50\%) in K band under ESO's median atmospheric profile, with 50\% sky-coverage at galactic pole.
In best-seeing condition MAORY is required to deliver 50\% Strehl ratio (goal 60\%) over MICADO's small field of 20x20" without sky-coverage conditions. Uniformity of the correction over the field has to be better of 10\% peak-to-valley, absolute Strehl values. 

As for the interfaces with ELT and instruments, MAORY makes use of the ELT's M4 5316-actuators deformable mirror \cite{2016SPIE.9909E..7YB} and of the telescope laser guide star facility.
The MCAO Natural Guide Star module (called LOR, Low Order and Reference WFS unit) has a dedicated volume in the field-derotated MICADO structure between the cryostat and the SCAO module in a gravity invariant orientation.   

A second output port for a future instrument located on the opposite side of the Nasmyth platform has to be provided, excluding the design of the NGS module.
A calibration unit made up of several modules for the daytime needs of both MAORY and MICADO is hosted at the entrance of the MAORY optical relay.     

The footprint of the MAORY volume on the Nasmyth platform is about 8x7m, with the input port being at 6000 mm above the platform and the MICADO port being at 4200 mm above the platform, gravity invariant and with 1800mm of free focal length; the mass budget for the instrument is 25t.  

\section{METHODS AND TOOLS TO ESTIMATE MCAO PERFORMANCES}

As for any other AO system, the performance of an MCAO system have a large variability depending on two aspects that are not fully under control of the observer: the atmospheric turbulence and the availability of natural guide stars. While in the SCAO case two scalar values (the integrated seeing value and the magnitude of the guide star) are typically used to fully parametrise the performances, in the MCAO case the vertical turbulence profile and the spatial distribution of the natural guide stars are also major determiners of the obtainable correction, making much harder the task of concisely describing the AO performances.

The need to develop a synthetic representation of MAORY's performances under a variety of observing conditions has driven the development of a suite of tools to provide a statistical estimate of the AO correction. 

\subsection{Estimating sky coverage}
A first step is the extension to a statistical representation of the sky-coverage, in which a given AO correction in the scientific field is expressed as a function of the fraction of sky providing NGS asterism capable of delivering the required AO correction, or better.
This statistical approach requires to simulate a large number of pointings in the sky and is described in detail in Ref.~\citenum{plantet2018wfs} and Ref.~\citenum{plantetAo4elt2019}.

In short, we realize a set of simulations using the end-to-end modeling code PASSATA \cite{10.1117/12.2233963}. Every simulation of the set is realized for the same given system configuration (same LGS asterism, WFS design, slope computing and control algorithms) and the same external conditions (atmospheric turbulence profiles, sodium layer, telescope elevation), but changing the NGS asterism for a total of 90 simulations (3 NGS asterisms, 6 magnitudes and 5 atmospheric realizations) that require 10-20h of computing time on standard GPU hardware. The LGS correction is the same for any given atmospheric realization, so it is saved once and restored to reduce computational time.

We then simulate 1000 random pointing in a field generated with TRILEGAL \cite{2005A&A...436..895G} mimicking the south galactic pole and for every pointing we select all the asterisms that are geometrically accessible to the MAORY NGS WFSs. For each asterism, the low order residual is computed as sum of windshake, tomographic error, noise and aliasing error that are estimated partly from analytical computation and partly as interpolation of end-to-end simulations results. 

The low order (LO) residual is then summed to the constant high order (HO) residual and to a fixed error budget containing estimates of errors due to NCPA, LGS elongation, telescope aberrations. The total wavefront error estimated in this way for each random pointing is used to compute the probability of having a Strehl ratio (using Marechal's approximation) equal or better than a given value. The graphical representation of the probability as a function of the Strehl ratio is referred in the following as "sky coverage curve"

\subsection{Selecting $C_n^2$ representative profiles}
\label{sec:cn2profiles}

In addition to the 5 contractually-binding "standard" atmospheric profiles provided by ESO for the instruments design, MAORY was welcomed to make best use of a large dataset \cite{2016SPIE.9909E..1NO,2017SPIE10425E..0BS} of 10000 $C_n^2$ profiles based on Stereo-Scidar measurement at Paranal. 
We decided to identify a method to order the $C_n^2$ profiles by MCAO performance and then select a few of them, namely the ones corresponding to 10, 25, 50, 75 and 90-th percentiles, and use them in the simulators to perform trade-off studies and guide the design of the AO system.

\begin{figure} [ht]
\begin{center}
\begin{tabular}[t]{c} 
\includegraphics[width=8cm]{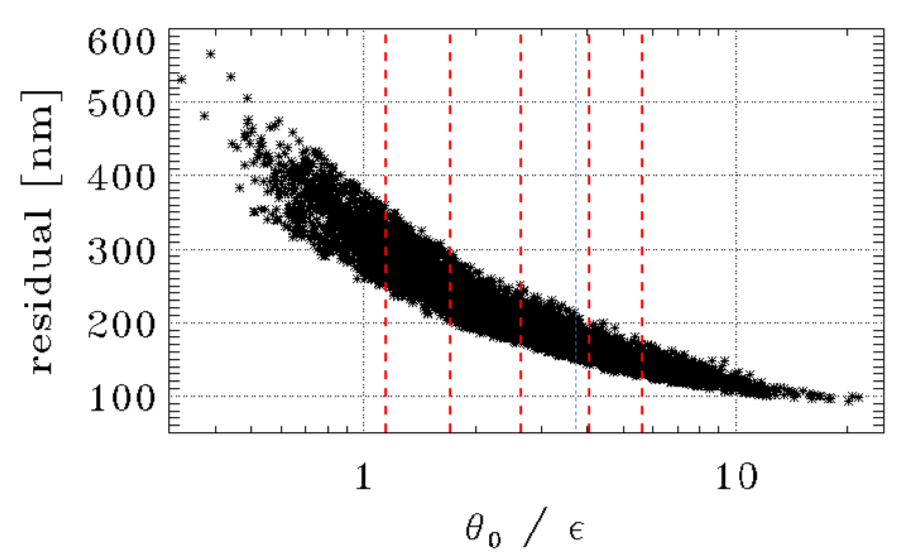}
\includegraphics[width=8cm]{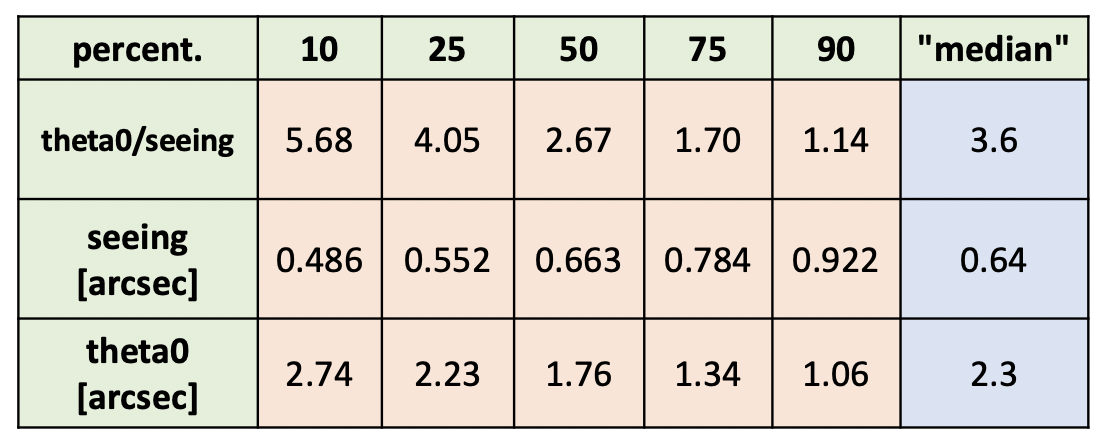}
\end{tabular}
\end{center}
\caption[example] 
   { \label{fig:cn2profiles} 
Expected high-order residual for the 10000 profiles of Ref.~\citenum{2017SPIE10425E..0BS} ordered as $\theta_0 / \epsilon$. The vertical red dashed lines correspond, from left to right, to $90^{th}$, $75^{th}$, $50^{th}$, $25^{th}$ and $10^{th}$ percentiles of $\theta_0 / \epsilon$. The table on the right summarizes seeing and isoplanatic angle values for the 5 profiles and for the original ESO "median" profile}
   \end{figure}

This approach presents two issues: a) the ranking according to the MCAO performances depends on the design of the AO system, so the ordering is not valid anymore when simulating a system with different parameters, and b) the size of the dataset requires to have a very fast tool, capable of assessing the expected performances for a given profile in a very short time.  
A fast Fourier analysis similar to the one described in Ref.~\citenum{2008JOSAA..26..219N} was used to evaluate on-axis residual for every profile and for a few system configurations and we found that the residuals show a decent correlation with the ratio between isoplanatic angle $\theta_0$ and seeing $\epsilon$ (see Fig.~\ref{fig:cn2profiles}, left).

So, in a purely empirical way, we decided to use the ratio $ \theta_0 / \epsilon $ to select the $i^{th}$ percentile profile, meaning that the probability of having a ratio $\theta_0 / \epsilon $ larger than the one of the chosen profile is $i$, and, because of the shown correlation, also the probability of having performances better than the one of the chosen profile is $\sim i$. 
In summary, by using this selection method, we identified the $50^{th}$ percentile profile that is representative of an atmospheric turbulence delivering median performance for any MAORY MCAO configuration.

A table summarizing $\epsilon$ and $\theta_0$ for the 5 selected profiles is reported in Fig.~\ref{fig:cn2profiles} (right), together with the corresponding values of the so-called "ESO median" profile defined in the MAORY Technical Specifications, that is $\sim$ 35 percentile according to the new dataset of profiles. 
To ensure that the design of MAORY is not biased toward too much favourable atmospheric conditions, in the mindset of following the rationale of the original requirements, we chose to evaluate MAORY performances using the 50-th percentile profile, named P50.

\section{Comparison of AO configurations}

Several comparisons studies have been carried out during the first period of phase B to determine the MAORY baseline AO configuration. In particular the analysis focused on the post-focal deformable mirrors specifications (number of DMs, conjugation altitude, number of actuators), on LGS WFSs (number of LGSs, asterism shape, number of subapertures and field of view) and on LOR WFSs (size of technical field, number of subapertures in LO and R WFSs, use of dedicated correctors in front of NGS WFSs \footnote{Similar to Dual Adaptive Optics proposed in Ref. \citenum{1992A&A...261..677R}})

Most of the above mentioned analysis have been addressed in another paper in this conference \cite{plantetAo4elt2019, obertiAo4elt2019} and we report here only the main findings: 
\begin{itemize}
    \item 3 DMs (M4 + 2 post focal) increases the performances by $\sim 5\%~SR_k$  compared to the 2 DM case
    \item the tomographic error is reduced by about 120nm when passing from 6 to 8 LGS at 45". Two-rings asterisms can also improve tomographic sensing.
    \item In the fast loop, NGS are used only for the Tip/Tilt slopes, to control TipTilt and $1^{st}$ order distortion modes. Still, the LO WFS requires 2x2 sampling because the focus detrending due to sodium layer altitude fluctuations requires a 10Hz framerate that would limit the sky-coverage if performed by the Reference WFS.
    \item A reduction of the technical field from 180" to 160" outer diameter has a marginal impact on sky-coverage,  and can be a acceptable trade-off if the optical design is facilitated. 
    \item The spot truncation effect can be mitigated with a proper tuning of the reconstruction algorithm, discarding part of the redundant slopes (see Ref.~\citenum{obertiAo4elt2019}).
\end{itemize}

In the following sections we focus on a few other items analyzed in the context of MAORY baseline definition.


\subsection{Post Focal DMs pitch}
We have studied the impact of post focal DMs pitch on the system performance.
In particular, we focused on two possible configurations: a pitch of 1.5-1.8m given by ~700 actuators DMs and a pitch of 0.6-0.7m given by ~4000 actuators DMs. The larger pitch corresponds to the case of voice-coil DM with actuator density of $\sim$ 25mm (as in the LBT ASM\cite{2008SPIE.7015E..12R} family) on a pupil size of about 800mm as in the current optical design. The smaller pitch corresponds to a reasonably high limit for the number of actuators of MAORY's DM in case a different technology is made available.

We set up simplified simulations, with perfect, noiseless and aliasing free WFSs and no turbulence evolution. In this way we are able to point out only the effects of tomographic and generalized fitting errors that depend on the DMs specifications. We replicate each simulation with 100 different atmospheric realizations to have a good statistical sampling.

\begin{figure}
    \begin{center}
    \begin{tabular}{c}
        \includegraphics[width=0.8\textwidth]{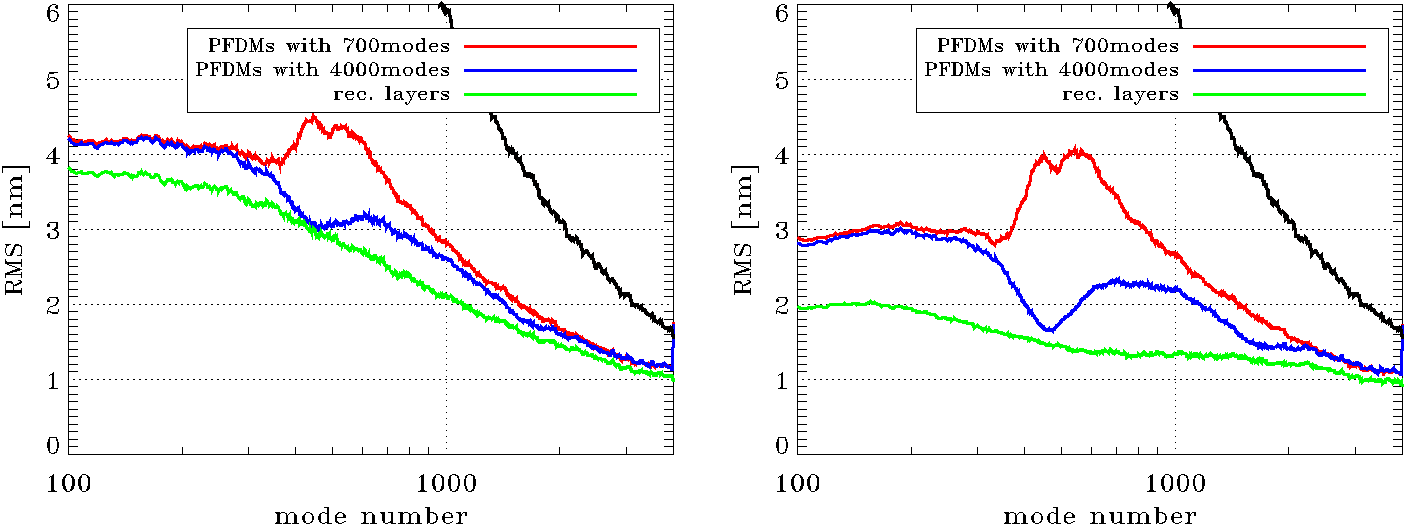}
    \end{tabular}
    \end{center}
	 \caption{\label{fig:pitch} On-axis tomography and generalized fitting modal error RMS, for a 6 (left) and 12 LGSs asterism (right). The green curve corresponds to the case of tomographic error only. For the sake of clarity a smoothing has been applied.}
\end{figure}

\begin{figure}
    \begin{center}
    \begin{tabular}{c}
        \includegraphics[width=0.8\textwidth]{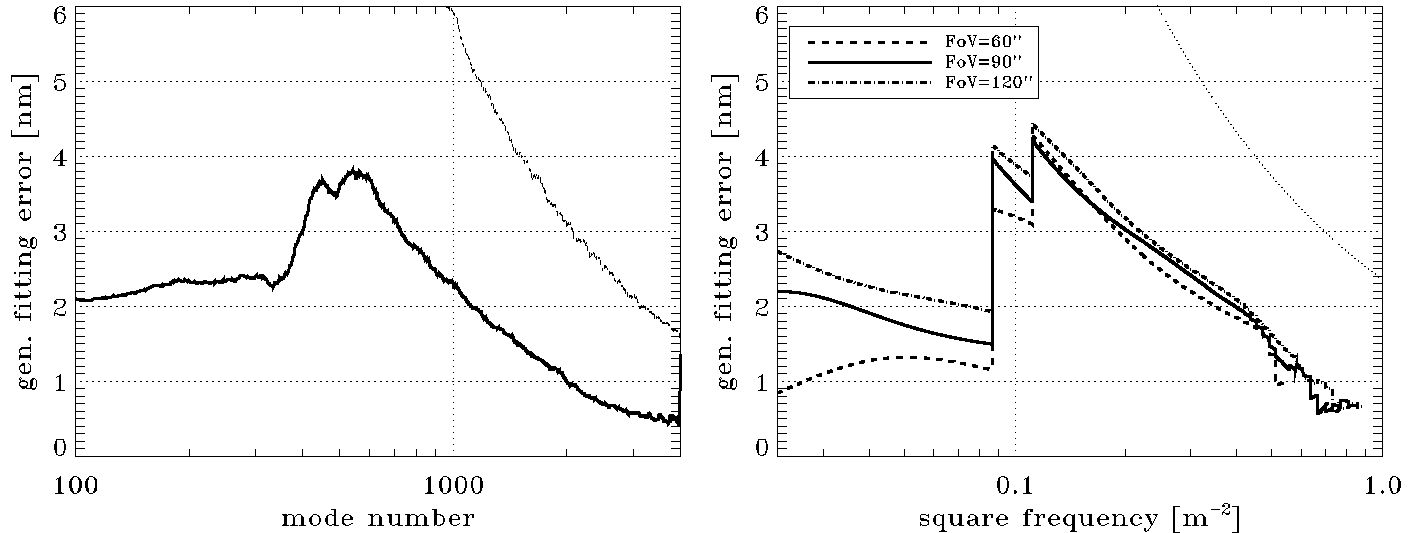}
    \end{tabular}
    \end{center}
	 \caption{\label{fig:genFit} On-axis generalized fitting error, computed as difference between red and green lines of Fig. \ref{fig:pitch} (left) and from analytical formulas\cite{10.1117/12.390311} (right). Different FoVs are considered to be consistent with end-to-end simulations.}
\end{figure}

The results are shown in Fig. \ref{fig:pitch} for 6 and 12 LGSs cases. We also compute the correction given by all reconstructed layers to have an estimate of the tomographic error only (green lines). One can notice that the difference between the two pitch is limited to modes in the range [300,2500]. Mode 300 approximately corresponds to a 1.8m pitch, that is the one of the highest (16km) and larger pitch DM, while mode 2500 corresponds to a 0.6m pitch, that is the one of the medium (8km) and smaller pitch DM. Moreover the difference on modes greater than 1000 is small because tomographic error (green line) is becoming the dominant error: only the 12LGSs asterism case is able to effectively exploit the advantage of the smallest pitch. We verified that the difference between red (or blue) and green curves is in good agreement with the theoretic generalized fitting error\cite{10.1117/12.390311} as can be seen in Fig.\ref{fig:genFit}.

\subsection{Number of LGS WFS subapertures}

We studied the impact of the number of sub-aperture in the LGS WFSs on the MAORY performance. We run end-to-end simulations for a system with 6 LGSs and favourable NGS asterism (bright stars close to the scientific field) under median atmospheric conditions; we then sampled the range between 40 and 80 subapertures along the diameter.

\begin{figure}
    \begin{center}
    \begin{tabular}{c}
        \includegraphics[width=0.8\textwidth]{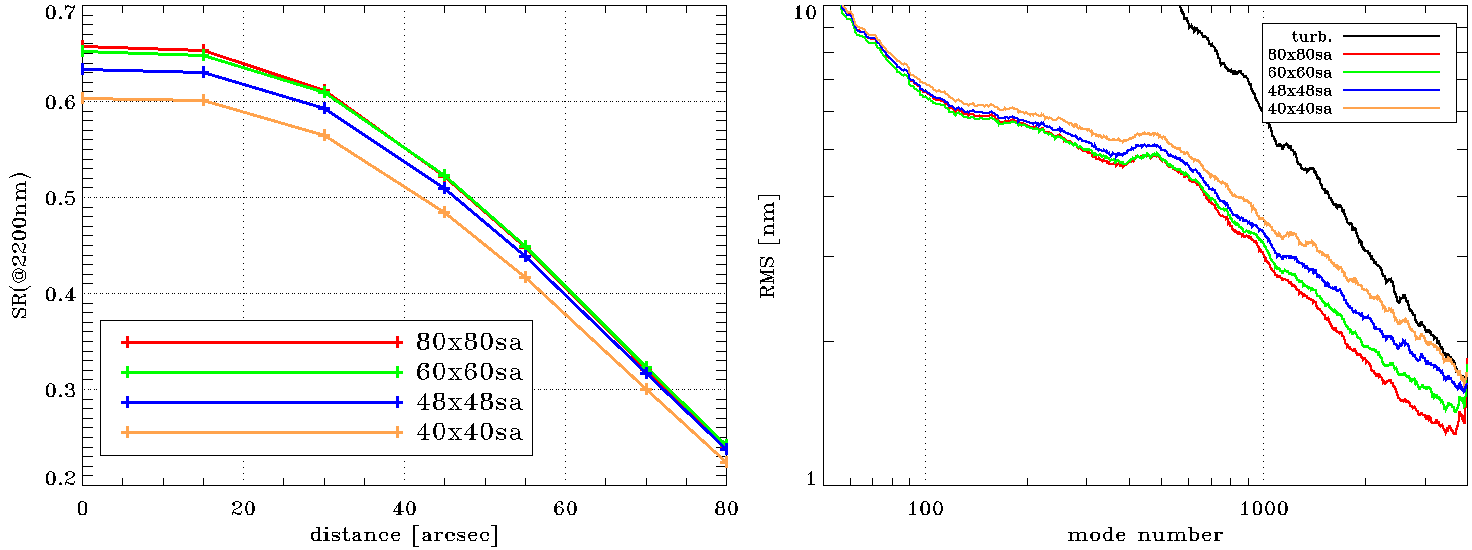}
    \end{tabular}
    \end{center}
	 \caption{\label{fig:nSaLgs} For different number of sub-apertures in the LGS WFSs: K band SR as a function of the off-axis distance (left) and modal residual (right)}
\end{figure}

The results are summarized in Fig.\ref{fig:nSaLgs}: on left part of this figure we can see that configurations with 80 and 60 sub-aperture give similar performances and the coarser sampling of 40 sub-aperture configuration has only 5\% less K band SR on axis.
All configurations are able to correct up to 4000modes as shown in the right part of Fig.\ref{fig:nSaLgs}.
We associate this result with the conjugation altitude of M4, that is optically conjugated at 620m and allows MAORY to take advantage of the different lines of sight of its LGS WFSs. Indeed, each WFS (considering the r=45" LGS asterism) sees a portion of M4 shifted by 0.131m with respect to the center: combining all the WFSs measurement, MAORY is able to properly sense all the actuators of M4 even with a reduced sampling of the individual LGS WFSs.
Only in the 40 subapertures case the correction is limited to about 3500modes.
Note that difference in performance mostly comes from the larger noise propagation and aliasing due to the lower sampling.
%

\subsection{Reference WFS specifications}

Reference WFSs in MAORY compensate errors in LGS WFS due to sodium layer (mismatched altitude conjugation and truncation effect) and any kind of non-common path aberrations (NCPA) between the LGS and the Reference WFS paths. 
These errors mainly impact on the low order modes: in particular the biases introduced by mismatched altitude and truncation effects are propagated as biases on focus and astigmatism only; NCPAs also rapidly decrease with spatial frequency.
Note that the focus correction will be based on LO WFSs as stated in Ref. \citenum{plantetAo4elt2019}.

\begin{figure}
    \begin{center}
    \begin{tabular}{c}
        \includegraphics[width=0.7\textwidth]{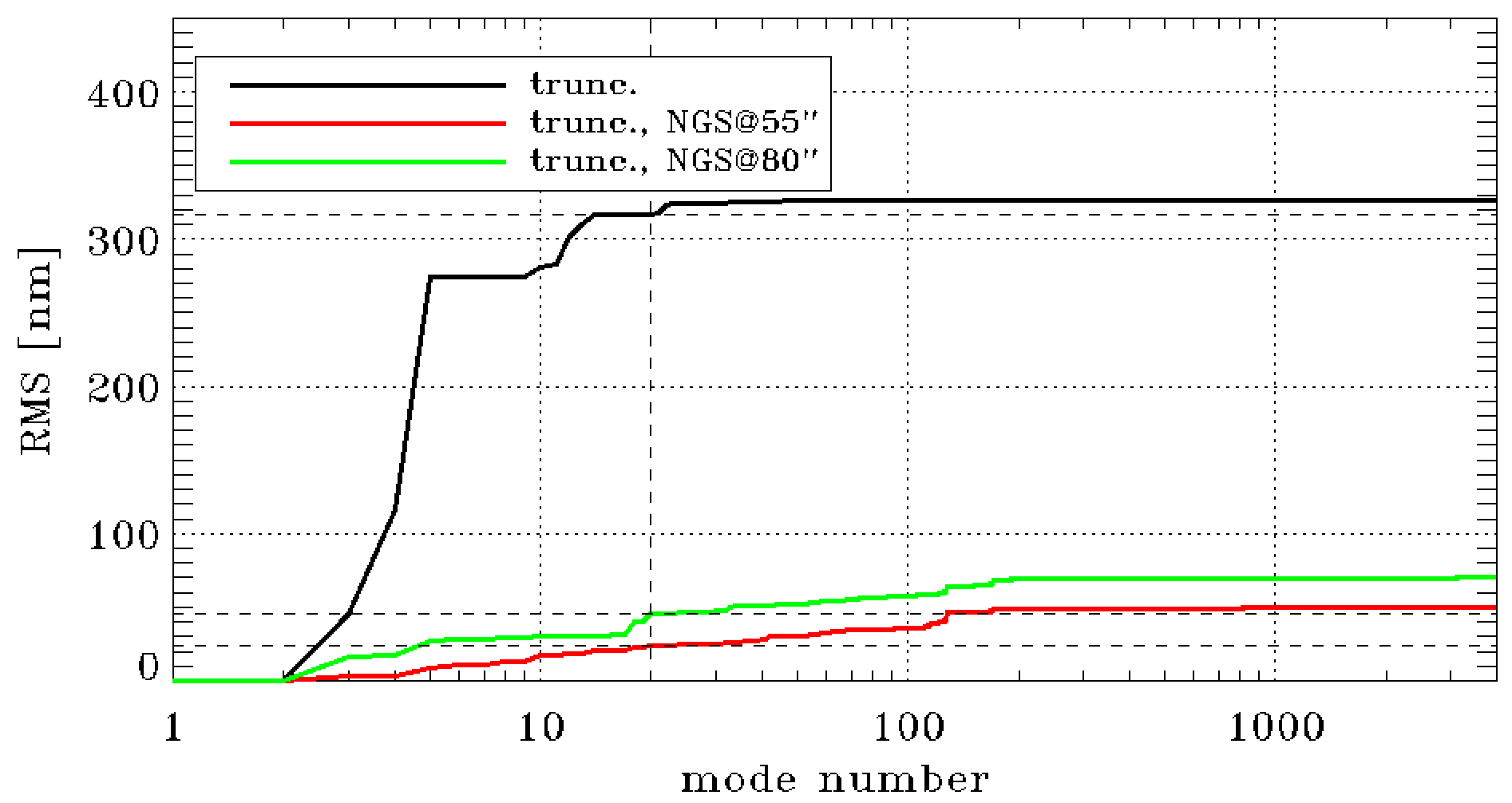}
    \end{tabular}
    \end{center}
	 \caption{\label{fig:multiPeak} Modal-cumulated truncation error on-axis from end-to-end simulation with the \emph{multi-peak} sodium profile \cite{2014A&A...565A.102P}, 10 arcsec FoV LGS WFSs, without atmosphere and without any reconstruction algorithm which takes into account spot elongation. Red and green curves consider a reference correction from NGSs at 55 and 80arcsec off-axis respectively.}
\end{figure}

We chose the \emph{multi-peak} sodium profile \cite{2014A&A...565A.102P} to evaluate the kind of errors that the reference WFSs must be capable to sense.
We found in end-to-end simulation that with 10 arcsec FoV LGS WFSs, without atmosphere and without any reconstruction algorithm which takes into account spot elongation (a pessimistic assumption only done for the purpose of this test) MAORY, on-axis, has an additional wavefront error of 330nm, but about 95\% of it is coming from the first 20 modes (see Fig.\ref{fig:multiPeak}).
Hence, we considered that a sampling of 10x10 sub-aperture is a reasonable first guess for the Reference WFS.

Then, we activated the reference loop with such 10x10 SHSs in the end-to-end simulation: this loop reduces truncation error to 50 and 70nm when NGS distance is 55arcsec and 80arcsec respectively (see Fig.\ref{fig:multiPeak}).

\begin{figure}
    \begin{center}
    \begin{tabular}{c}
        \includegraphics[width=0.7\textwidth]{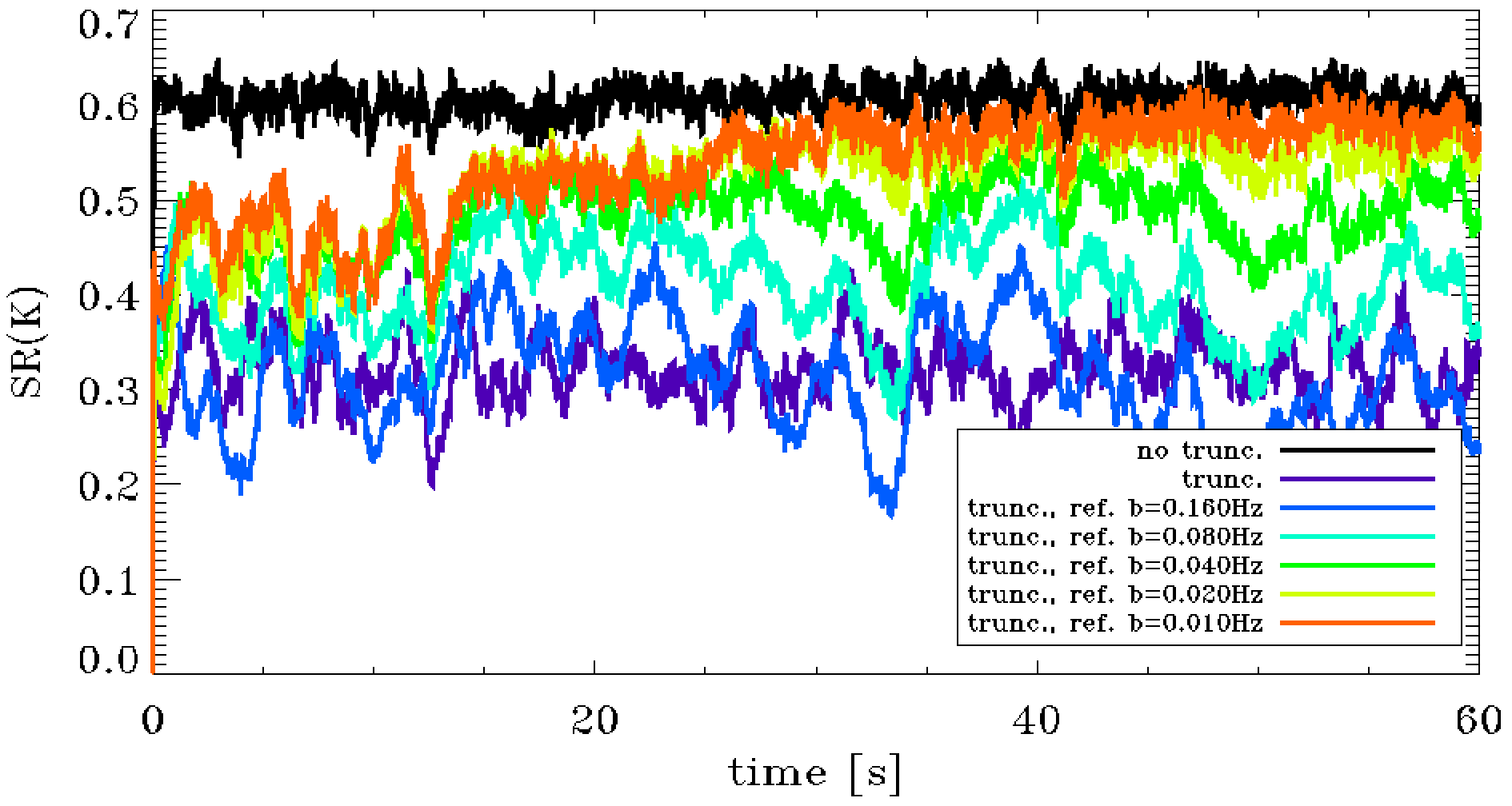}
    \end{tabular}
    \end{center}
	 \caption{\label{fig:srRefBandwidth} K band SR as a function of the time for different bandwidths of the reference control.}
\end{figure}

Finally, we tested the efficiency of reference loop when atmospheric turbulence is present. In this case we tried several control bandwidths and we saw that the reference loop efficiency increases with smaller ones (see Fig. \ref{fig:srRefBandwidth}). The smallest additional error amounts to 60nm and it comes from a 0.01Hz bandwidth. 

It must be considered that the reference loop is not able to discriminate between turbulence and static errors; by decreasing the control bandwidth we are able to avoid that the reference control loop starts acting on the  correction of the turbulence that is unefficient because the reference WFSs have a significantly larger tomographic error than the LGS WFSs, since they are only 3 WFSs arranged on wider asterism ($>$ 55 arcsec radius) than the LGSs's one (45 arcsec radius).

As a final note, using the small bandwidth described above for the Reference control loop allow us to work with the faintest NGSs that contribute to the sky coverage of the system: H=22 and R+I=24.


\subsection{Control algorithms}
\label{sec:control}

MAORY will use the information coming from the LGS and LOR WFSs to make a tomographic wavefront reconstruction and control.
The control architecture is based on some well known algorithms: tomographic recontruction and DM projection
as detailed in Ref.~\citenum{Fusco:01}, pseudo-open loop control (POLC) as detailed in Ref.~\citenum{10.1117/12.506580} 
and split tomography as detailed in Ref.~\citenum{Gilles:08}.

Thanks to split tomography, HO, LO and Reference/truth sensing (R) are reconstructed independently by laser, natural IR and visible WFSs respectively, as: 
\begin{equation}\label{eq:polc}
    \Delta c(k) = P W s(z) - H c(k-d)
\end{equation}
where $c$ is the DMs modal command vector, $\Delta c$ is the incremental DMs modal command vector, $P$ is the projection matrix from reconstructed layers modes to DMs modes, $W$ is the tomographic reconstruction matrix from WFSs slopes to reconstructed layers modes, $s$ is the WFSs slope vector, $k$ is the time step, $d$ is the total loop delay in frames and $z$ is the variable of the Z-transform. 
$H$ is a square matrix defined as:
\begin{equation}\label{eq:H}
    H = I - P W D_{DM}
\end{equation}
where $I$ is the identity matrix and $D_{DM}$ is the interaction matrix from DMs modes to WFSs slopes.

Note that we introduced a few minor differences with respect to the classic POLC algorithm:
time filtering is done in the DMs modal space and the pseudo-open loop slopes are not explicitly computed.
These choices, which have a negligible impact on performance, were made to reduce the computational complexity.

We spend an additional sentence to point out the meaning of the $H$ matrix:
it gives the expected value of the measurement noise propagated in the closed loop and, hence, it avoids unseen or badly seen modes to propagate further in the closed loop.
To show this and for reason of simplicity, we consider the case where the reconstructed layers correspond to the DMs ($P=I$ and $D=D_{DM}$).
Recalling that $W$ is a minimum-mean-square-error estimator \cite{Fusco:01} we find that:
\begin{equation}\label{eq:H_derivation}
    H = I - W D = I-(D^T  C_N^{-1} D + C_{\Phi}^{-1} )^{-1} D^T C_N^{-1} D = (D^T  C_N^{-1} D + C_{\Phi}^{-1} )^{-1} C_{\Phi}^{-1}
\end{equation}
where $C_N$ and $C_{\Phi}$ are the noise and turbulence covariance matrices respectively.

After HO, LO and R incremental command vectors are available, the final DMs command vector is computed as:
\begin{equation}\label{eq:timeFilter}
    c(k) = G(z) \left( \Delta c_{LHO}(k) + c_R(k) \right)
\end{equation}
\begin{equation}\label{eq:timeFilterReference}
    c_R(k) = G_{R}(z) \Delta c_{R}(k)
\end{equation}
where $G$ is the temporal filter, $\Delta c_{LHO}$ is the incremental DMs command vector given by the combination of LO and HO modes and $c_R$ is the reference modal offset. We chose to implement a modal offset instead of a slope offset to avoid an additional MVM.

\section{MAORY adaptive optics baseline configuration}

In short, the main findings from AO analysis to be used at the level of system design and subsystem specifications are the followings:
\begin{enumerate}
\setcounter{enumi}{0}
\item two post-focal DMs with 700+ modes are needed to reach the  30\% SR in K band at 50\% sky-coverage with P50 atmospheric profile.
\item At least 6 LGS are needed to fulfill the requirements. 4 LGS are not enough.
\item ultimately,  to  increase  the  performance,  the  system  is  limited  by  the  tomographic reconstruction  so  by  the  number  of  laser  guide  stars:  with  8  LGS  the  performances could improve by 5-10\% of SR in K (depending on  atmospheric profile). In this case the system will require DMs with 1500+ actuators.
\item the fast NGS WFS are required to sense only tip-tilt (although a 2x2 SH is still needed for  fast  focus  detrending):  this  control  scheme  improve  significantly  the  sky coverage over the previous one based on 5 modes sensed on NGS WFS.
\item the spatial sampling of LGS WFS can be reduced from 80 to about 60 subapertures without significant performance loss. This could help to obtain a larger field for the LGS subapertures and reduce the impact of sodium spot truncation.  
\end{enumerate}

\begin{figure}
    \begin{center}
    \begin{tabular}{c}
        \includegraphics[width=0.45\textwidth]{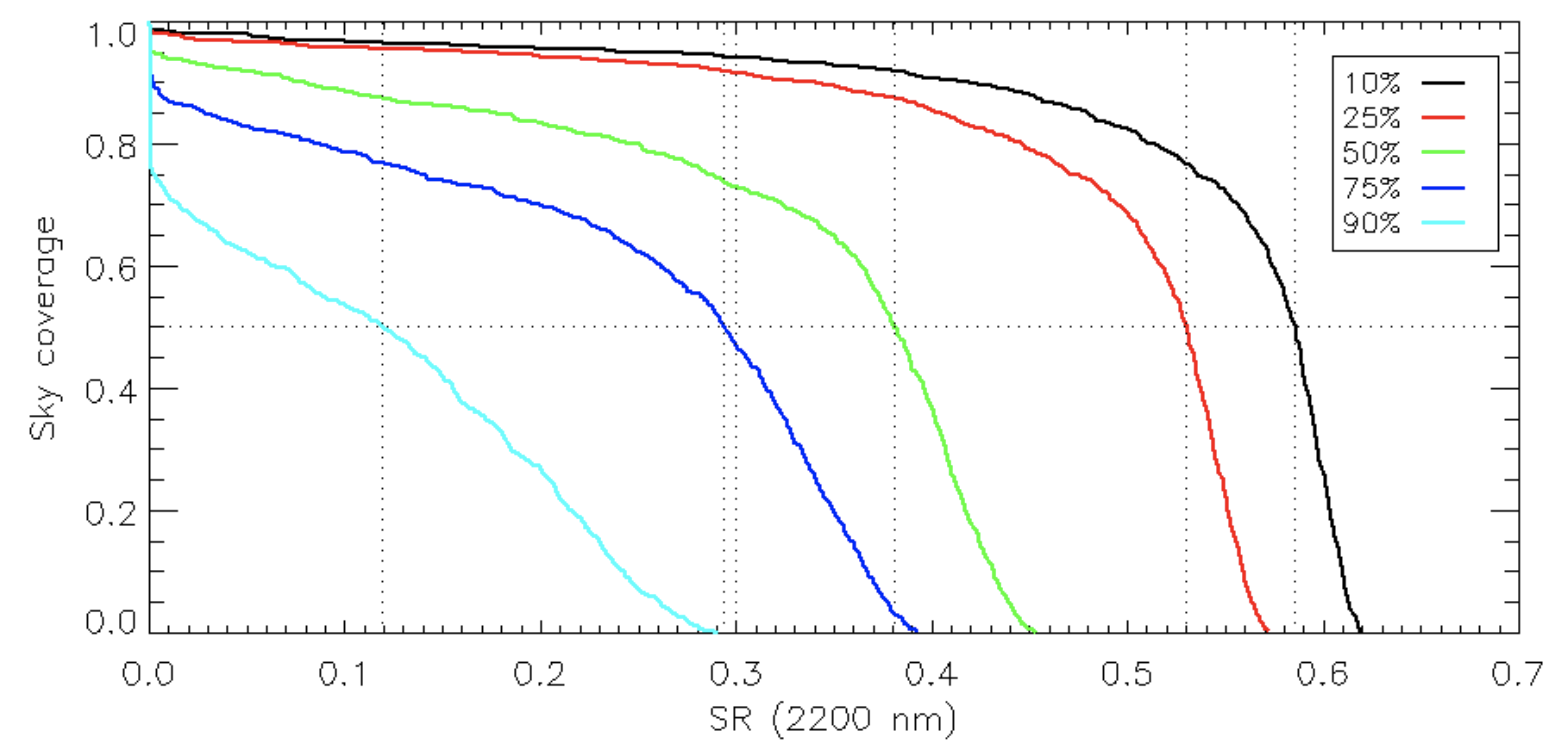}
        \includegraphics[width=0.45\textwidth]{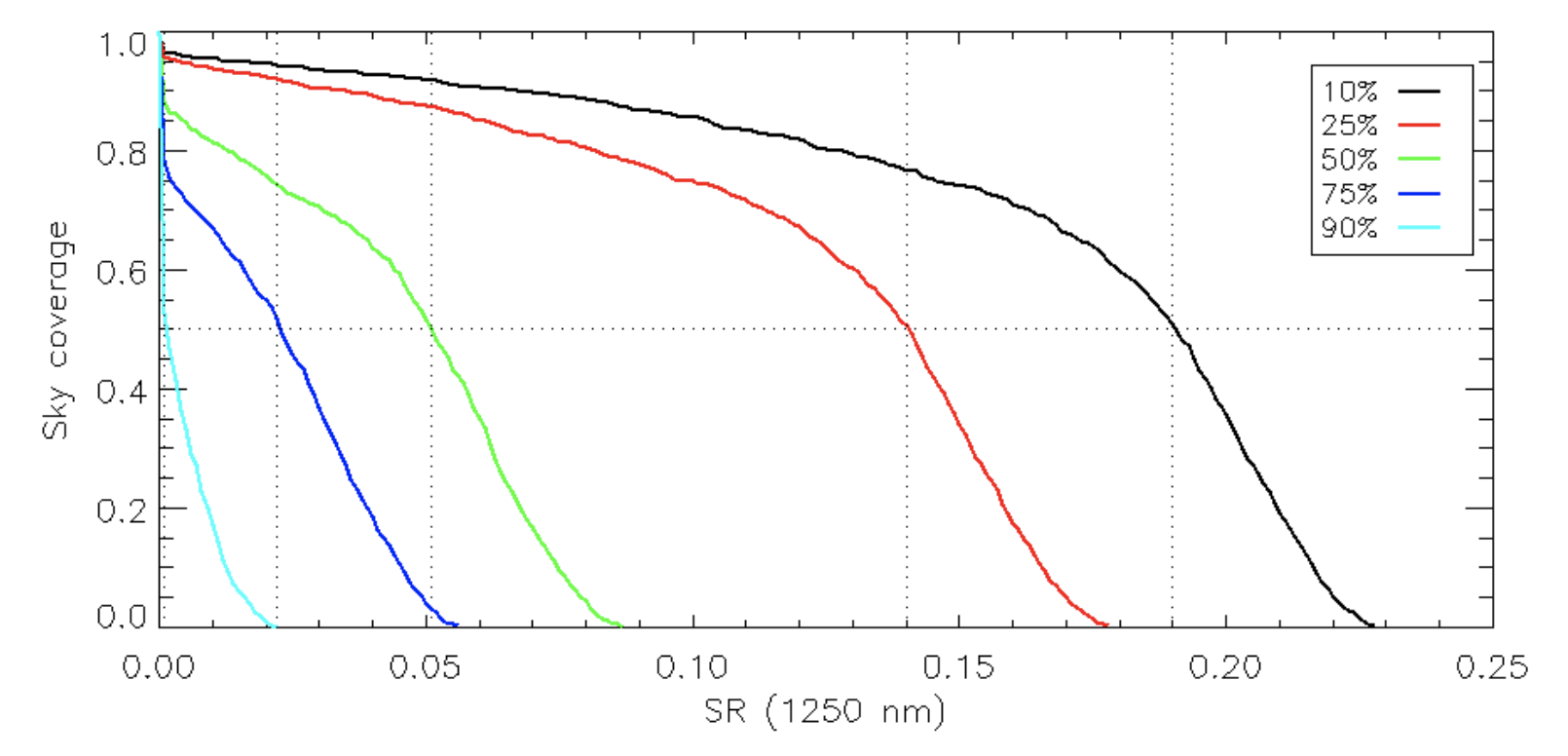}
    \end{tabular}
    \end{center}
	 \caption{\label{fig:skyCoverage} Sky coverage curve for different atmospheric conditions as defined in Sec \ref{sec:cn2profiles}. Strehl ratio in K band (left) and J band (right) for an AO system based on 2 post focal DMs correcting 1500 modes each, 8 LGSs at 45”, 30$^{\circ}$ zenith angle  }
\end{figure}

\begin{figure}
    \begin{center}
    \begin{tabular}{c}
        \includegraphics[width=0.7\textwidth]{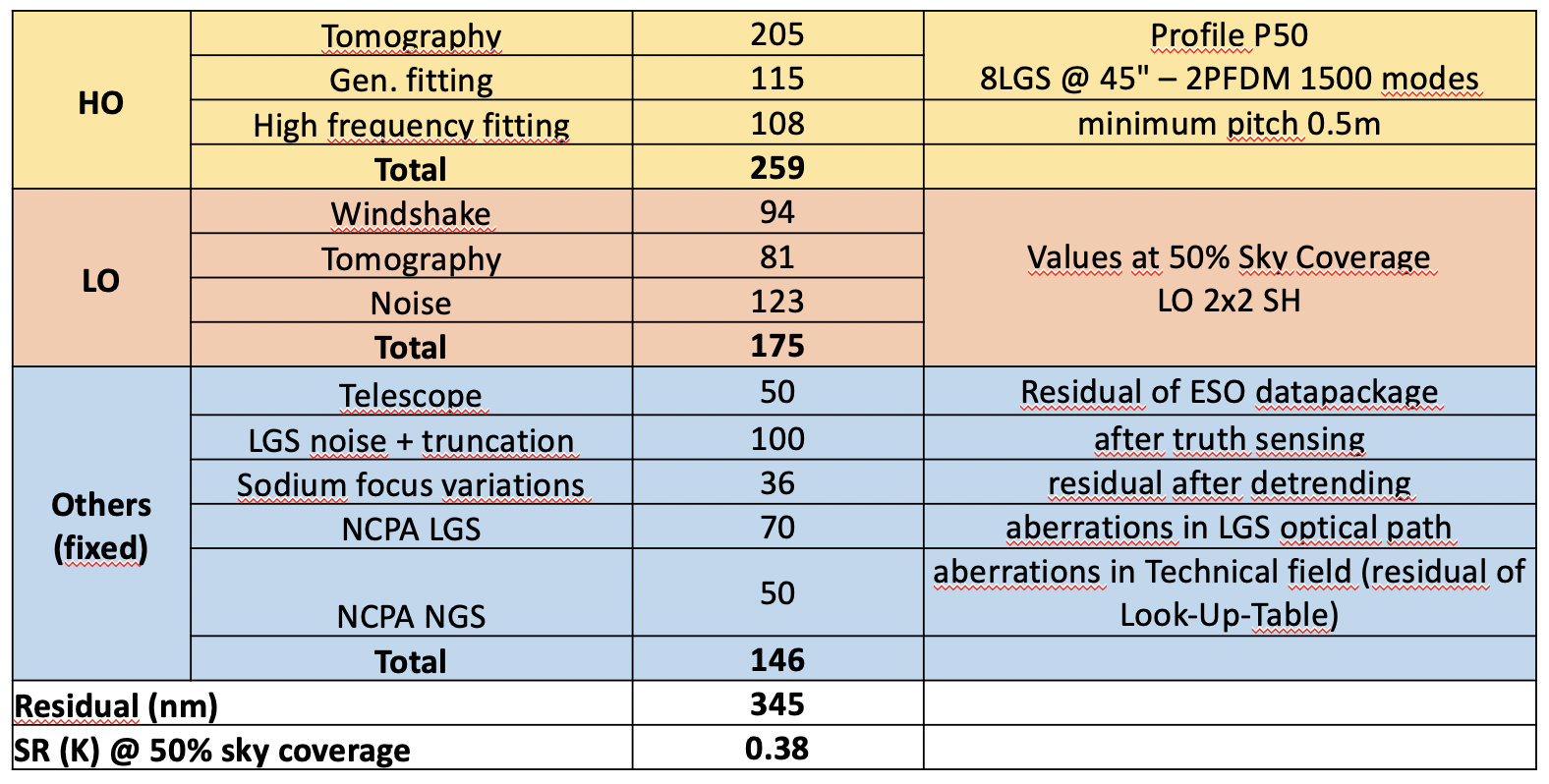}
    \end{tabular}
    \end{center}
	 \caption{\label{fig:errorBudget} Error budget for the baseline system described in the text based on 2 post focal DMs correcting 1500 modes each, 8 LGSs at 45”, evaluated for the P50 atmospheric profile, at 30$^{\circ}$ zenith angle  }
\end{figure}

The final AO configuration will be chosen taking into account several other constraints coming from the other components of MAORY, from the ELT and from the interfaces with the surrounding instruments and of course from considerations of schedule and cost. 

The current AO baseline configuration consists in 3 DMs with the 2 post focal DMs having more than 1500 actuators each, 8 LGS WFS arranged on a regular asterism of 45" radius with a 15" FoV, 3 LORs patrolling a 160" technical field, 2x2 LO WFS and control scheme as described in Sec.~\ref{sec:control}. The performances for such a system configuration are shown in Fig.~\ref{fig:skyCoverage}, and the error budget for the P50 profile is shown in Fig.~\ref{fig:errorBudget}.

\bibliography{report} 
\bibliographystyle{spiebib} 
\end{document}